# Atomistic Evidence of Nucleation Mechanism for the Direct Graphite-to-Diamond Transformation

Duan Luo[1,2,3†], Liuxiang Yang[4†], Hongxian Xie[5], Srilok Srinivasan[1], Jinshou Tian[2], Subramanian Sankaranarayanan[1], Ilke Arslan[1], Wenge Yang[4*], Ho-kwang Mao[4,6*], Jianguo Wen[1*]

[1]Center for Nanoscale Materials, Argonne National Laboratory, Lemont, Illinois 60439, USA

[2]Key Laboratory of Ultra-fast Photoelectric Diagnostics Technology, Xi'an Institute of Optics and Precision Mechanics, Chinese Academy of Sciences, Xi'an 710119, China.

[3]University of Chinese Academy of Sciences, Beijing 100049, China.

[4]Center for High Pressure Science and Technology Advanced Research, Beijing, 100094, China

[5]Hebei University of Technology, Tianjin 300132, China

[6]Geophysical Laboratory, Carnegie Institution of Washington, Washington, DC, 20015, USA

*Corresponding author. Email: yangwg@hpstar.ac.cn (W.G.Y.), maohk@hpstar.ac.cn (H.K. M.), jwen@anl.gov (J.G.W.)

†D.L. and L.Y. contributed equally to this work.

**The direct graphite-to-diamond transformation mechanism has been a subject of intense study and remains debated concerning the initial stages of the conversion, the intermediate phases, and their transformation pathways. Here, we successfully recover samples at early conversion stage by tuning high-pressure/high-temperature conditions and reveal direct evidence supporting the nucleation-growth mechanism. Atomistic observations show that intermediate orthorhombic graphite phase mediates the growth of diamond nuclei. Furthermore, we observe that quenchable orthorhombic and rhombohedra graphite are stabilized in buckled graphite at lower temperatures. These intermediate phases are further converted into hexagonal and cubic diamond at higher temperatures following energetically favorable pathways in the order: graphite → orthorhombic graphite → hexagonal diamond, graphite → orthorhombic graphite → cubic diamond, graphite → rhombohedra graphite → cubic diamond. These results significantly improve our understanding of the transformation mechanism, enabling the synthesis of different high-quality forms of diamond from graphite.**

Direct diamond formation from graphite without catalysts[1, 2, 3, 4] has attracted immense attention for more than sixty years because diamond has the highest hardness and thermal conductivity of any natural material[5, 6, 7]. Synthesis of diamond has been successfully accomplished by two main approaches: static compression[2, 3, 4] or shock compression of hexagonal graphite (HG)[1, 8, 9]. Under static compression larger than ~9-18 GPa at room temperature, HG converts into a transparent high-pressure carbon phase[10, 11, 12, 13, 14], but it reverts back to HG after pressure release. Further heat treatment exceeding ~1073 K[3, 14, 15, 16] under high pressure can synthesize both hexagonal diamond (HD) and cubic diamond (CD). The formation of HD is favored at relatively low temperature[3, 15], while higher temperatures (especially > ~1873 K[16]) increase the yield of CD[2, 4, 15, 16, 17]. Although significant efforts have been devoted to formation of diamond from graphite, the underlying microscopic mechanism for the G-D transformation is still under debate, particularly concerning the initial stages of the conversion[18, 19, 20], the intermediate metastable phases[21, 22, 23], and their transformation pathways[18, 19, 20, 21, 22, 23, 24, 25, 26, 27].

Due to the lack of matching carbon atoms at the honeycomb centers with carbon atoms in adjacent graphitic layers, extensive theoretical researches predict that the direct G-D transformation must first undergo a sliding process to form intermediate phases. Two intermediate phases, orthorhombic graphite (OG) with AB' stacking[21, 28] and rhombohedral graphite (RG) with ABC stacking[22, 29], are identified to bridge the direct transformation through boat-buckling or chair-buckling, respectively (Fig. S1). Although numerous experiments[3, 18, 30] supported these concerted pathways by the crystallographic orientation relationship between graphite and diamond (HD/CD), there has never been direct atomistic observation of these transformation pathways. Furthermore, the energetic barriers for these concerted transformations are too high for the homogeneous diamond growth[19, 20, 31].

Recently, Khaliullin et al.[19] proposed an energetically more favorable and more realistic mechanism for diamond formation where G-D transformation begins from a diamond nucleus followed by sequential growth into a graphite matrix using a neural network potential. This theoretical study showed that large lattice distortions due to the formation of diamond nuclei inhibit the phase transition at low pressure and direct it towards the HD phase at higher pressure[19]. Furthermore, using stochastic potential energy surface global exploration, Xie et al.[26] found that HD has a facile initial nucleation mechanism and faster propagation kinetics over CD (> 40 times) in graphite matrix because of three low-energy coherent graphite/HD interfaces. However, to date, the existence of the hypothesized intermediate phases and the formation of diamond nuclei in the G-D transformation have never been experimentally observed at the atomic scale. This is mainly due to the lack of partially converted samples, which allows direct atomistic observation of nucleation and growth of diamond from graphite. Although Garvie et al[32] reported a topotactic mechanism by studying an incompletely transformed diamond in an extraterrestrially shocked meteorite composed of graphite, it is naturally harvested and cannot be repeated unlike synthesized samples.

Here, we report direct evidence of the nucleation-growth mechanism and the energetically favorable transformation pathways of the direct G-D using high-resolution transmission electron microscopy (HRTEM). Atomistic observations show that metastable OG phases mediate the growth of nanodiamond nuclei from the graphite matrix. By tuning HPHT, we also obtain quenchable OG located between OR/OR or OR/HG regions. Furthermore, we find both HD and CD are converted from HG via sliding of graphite planes into metastable OG under mild HPHT conditions, and then through boat-buckling or chair-buckling respectively. Combining these atomic-resolution observations with large-scale MD simulations and DFT calculations, we have been able to identify a diffusionless nucleation-growth mechanism through HG to OG, then to HD or CD.

**Nucleation from Hexagonal Graphite**

Single-crystal graphite disk samples were loaded into a diamond anvil cell (DAC) compressed to ~20 GPa with silicone oil or neon serving as a pressure medium, and then heated with a YAG laser to different temperatures (1000 K, 1200 K, 1400 K). After quenching and releasing pressure, we found the sample heated at 1400 K is transparent while the other two samples are opaque. As reported in our previous paper[33], the transparent sample synthesized at 1400 K is completely converted into a HD phase by the HPHT synthesis. X-ray diffraction (XRD) combined with HRTEM show that the recovered HD sample is a high-quality single phase (Fig. S2) with the crystallographic orientation relationship of $(001)_{HG}$ // $(100)_{HD}$ and $(100)_{HG}$ // $(001)_{HD}$. The opaque sample synthesized at 1000 K reverts back to graphite and the sample synthesized at 1200 K shows an incomplete transformation from graphite to HD.

Unlike complete and non-G-D transition samples, the incomplete transition sample is ideal for studying the nucleation mechanism and intermediate phases. TEM analysis reveals that the 1200 K sample contains substantial amounts of nanoscale HD embedded in the graphite matrix (Fig. 1a and 1b). Nanobeam electron diffraction and fast Fourier transform (FFT) patterns (Fig. S3c) from the HD (Fig. 1b) indicate that the 6-fold symmetry diffraction pattern has a 0.218 nm lattice spacing as indicated by a circle. This diffraction pattern can be only indexed as [001] diffraction pattern of a HD structure ($P6_3/mmc$, a= 2.51 Å, c= 4.16 Å) as discussed in Fig. S3. Electron energy-loss spectra (EELS) from these HD nuclei show similar fine K-edge structures as CD, confirming that these nanoscale HD have a $sp^3$ bonded diamond structure since to the peak at ~283 eV corresponding to $sp^2$ bonds are absent in Fig. S3g. The observation of these nanoscale HD further supports that HD phase is kinetically favored over CD under mild HPHT conditions[26].

The size of these nanodiamonds (Fig. 1a-1c) ranges from a few nm to ~20 nm. The experimental observations of these nanodiamonds in the graphite matrix directly indicate that the G-D transformation goes through a nucleation mechanism instead of the concerted mechanism as predicted by previous theoretical simulations[19, 26]. The smallest dimension of the nanoscale HD observed is about 2.5 nm in Fig. 1a, which is close to the predicted critical in-layer nucleation size in the simulation[19]. The shape of each individual nanodiamond nucleus varies from elongated (Fig.

1a), regular (Fig. 1b), to irregular hexagonal shapes (Fig. 1c), but each is enclosed by three pairs of {100} facets.

**Atomistic Evidence of Direct G-D Transformation via Intermediate Phase**

To gain further understanding of the direct G-D transformation pathways, we investigate the atomic structures at the HG/HD interface to verify whether there exist intermediate phases such as OG and RG. In Fig. 1b and 1c (details in Fig. S3-S5), we observe an intermediate OG phase next to nanoscale HD, indicating that OG mediates the growth of HD during the direct G-D transformation. Although FFT patterns (Fig. S3-S4) from these OG regions support that they are OG phases instead of HG, the best orientation to distinguish HG (AB stacking), OG (AB' stacking), and RG (ABC stacking) from each other is to view along $[100]_{HG}$ instead of along $[1\bar{1}0]_{HG}$ (or $[001]_{HD}$ in Fig. 1). We prepared TEM specimen orientated along $[100]_{HG}$ using the focused-ion beam. We find that dominant interfaces are graphite/HD and only a few of them are graphite/CD, consistent with the favorable growth of HD under mild HPHT conditions.

Fig. 2 shows HRTEM images for the graphite/HD and graphite/CD interfaces, revealing intermediate phases next to the nanodiamond nuclei. This intermediate phase is nearly identical to HG with an interlayer spacing of ~0.34 nm and in-layer spacing of 0.214 nm, but the stacking sequence is different from HG with AB stacking. In these HRTEM images, two in-layer carbon atoms are 0.08 nm apart and are imaged as one dot. These dots are aligned right on top each other between adjacent layers, rather than shifting half-way between layers for the AB-stacked HG. This intermediate phase with aligned dots matches with OG with AB' stacking or graphite with AA stacking (Fig. S5). By tilting the sample, we confirm that the intermediate phase is OG instead of graphite with AA stacking (Fig. S5). Fig. 2c and 2f show the orientation relationships between OG/HD and OG/CD as follow: $[100]_{OG}$ // $[100]_{HD}$ and $[100]_{OG}$ // $[110]_{CD}$. When the transformation goes through the HG→OG→HD/CD pathway, the graphite basal planes are expected to be perpendicular to the stacking planes $(001)_{HD}$ or $(111)_{CD}$ (Fig. S1). If the transformation goes through the transformation HG→RG→CD pathway, the graphite basal planes are expected to be parallel to $(111)_{CD}$ (Fig. S1). In both Fig. 2c and 2f, we observe that the graphite basal planes are perpendicular to the stacking planes $(001)_{HD}$ or $(111)_{CD}$, further indicating the transformation HG→OG→HD/CD pathways instead of HG→RG→CD pathway[21]. These relationships indicate that both HD and CD are converted from graphite via intermediate OG through boat- or chair-buckling at mild conditions, respectively.

**Formation of the Intermediate Phase**

It is crucial to understand how OG is formed from HG prior to the transformation into HD/CD. To address this issue, we lower synthesis temperature (1000 K) to obtain quenchable intermediate phases. Low-magnification TEM images show the recovered samples revert to graphite but consist of many voids with a rhombus shape (Fig. 3a). These voids are formed by buckling of graphite basal planes due to the anisotropic compression of high pressure and fluctuations of high temperature[3,4,19]. Within each rhombus, quenchable OG and RG phases are observed in the bent areas labeled in Fig. 3b. OG phases are always formed together with RG or HG (Fig. 3c) and no isolated orthorhombic crystallites have been detected, indicating that OG is hard to be quenched unless stabilized by RG, HG, or diamond. This can explain why metastable OG has never been reported, unlike RG which can be found in nature.

In the HPHT samples synthesized at 1200 K, nanoscale HD are often found in the buckled graphite regions next to voids (Fig. 1c). These areas coincidently correspond to the OG/RG regions in the 1000 K samples as schematically shown in Fig. 3b. Since the transformation from RG to HD has not been reported in any of the previous experimental studies, HD nucleation is therefore likely to initiate at these OG regions. In addition, the distortion of graphite basal planes creates planar defects at the interfaces[34] between OG and RG or HG. These structural defects might also play an important role in the direct G-D transformation[19].

To gain further insight into the atomistic mechanism of the G-D transformation, DFT calculations and large-scale MD simulations were performed (see Methods). Fig. S6 shows the enthalpy landscape for the concerted transformation HG→OG→HD pathway. The calculated activation energy for HG→OG →HD transition is ~0.3 eV/atom, which indicates the concerted mechanism is not favorable at these mild HPHT conditions. The nucleation process for the direct G-D transformation was investigated by MD simulations using a large initial HG model size (132 Å ×102 Å ×136 Å in the $[210]_{HG}$, $[100]_{HG}$ and $[001]_{HG}$ directions, containing 192,000 atoms). As shown in Fig. 4a, HD nuclei can only be stabilized over a critical nuclei size due to the extremely high surface energy of diamond[19]. Nucleation and growth happened at different areas with different time and speed. High pressure brings the graphite layers much closer together, at the same time causing shifts in favor of the OG and RG[29]. In the nucleation-growth process, the graphite basal planes undergo local 'in-layer' distortions that bring the atoms of the surrounding graphite matrix into an AB' stacking sequence (Fig. 4b-4e). At the beginning, only HG phase exists in the simulation model (Fig. 4b). When the pressure increases to a critical value, HD begins to nucleate at the top left corner of the model (Fig.4c); then grows rapidly via boat-buckling of OG, which is converted from HG phase (Fig 4d). After the pressure release, Fig. 4e shows clearly that OG mediates between HG matrix and HD, which is consist with the experimental results very well (Fig.1). Although there are some differences in the experimental and simulation conditions, MD simulation is qualitatively highly consistent with the experimental observations. The G-D transformation is governed by a diffusionless nucleation-growth mechanism under mild pressures (<20 GPa). And the intermediate OG phase plays a key role in the direct G-D transformation.

**Discussion**

The present results clearly reveal that the diffusionless nucleation mechanism is favored over the concerted transformation under mild HPHT conditions. Following the HG→OG→HD/CD pathways, OG can be converted into either HD or CD through boat-buckling or chair-buckling, respectively. Experimental observations indicate that HD nucleation is favorable over CD under mild HPHT conditions, although both pathways via OG are available. Such pathways also exclude the possibility of HD as the intermediate phase between HG and CD transition[8, 35]. Moreover, despite we observed small amount of CD, the transformation pathway HG→OG→HD is more energetically favorable than HG→OG→CD at mild HPHT conditions. Recent experimental results[33] showed the successful synthesis of single-phase HD crystal recovered from HPHT, further indicating that CD nucleation is suppressed under these synthesis conditions and HD exists as a discrete material rather than just a faulted or twined diamond[36,37]. In addition, HD is predicted to 58% harder than CD[38]. However, it's quite difficult to reliably verify this prediction because no single-crystal HD has been successfully synthesized up to now. Understanding the direct G-D nucleation mechanism and the corresponding transformation pathway guides us towards the possible routes to synthesize single-crystal HD. A high-quality HD may also provide a new platform for hosting nitrogen-vacancy centers for quantum information since HD has a different band structure with CD.

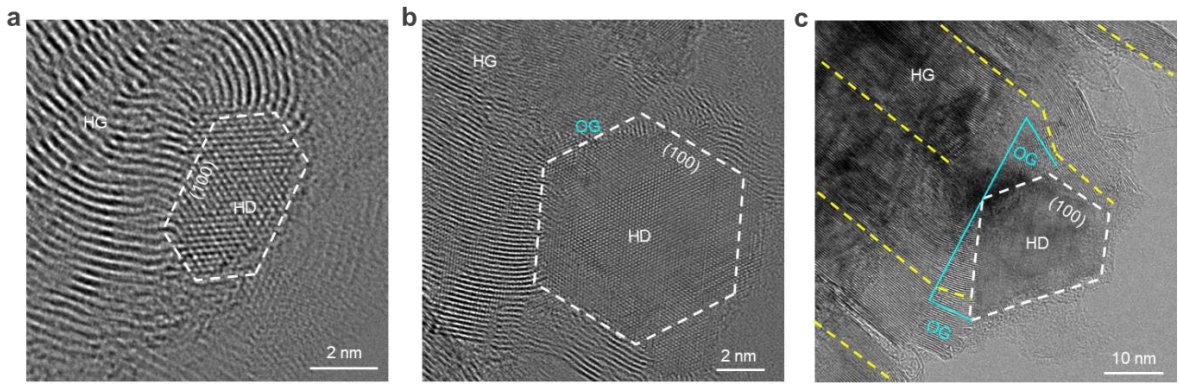

**Fig. 1 Nanoscale HD nucleation and growth mediated by the intermediate OG phase**. **a-c**, HRTEM images of nanoscale HD nuclei with different sizes and shapes embedded in the graphite matrix. OG phase is clearly observed next to HD in **b** and **c**. The FFT patterns from these HDs can be indexed by $[001]_{HD}$. The FFT patterns from the OG in **b** and **c** show a diffraction pattern matching with $[310]_{OG}$ instead of $[1\bar{1}0]_{HG}$ (Fig. S3-S5).

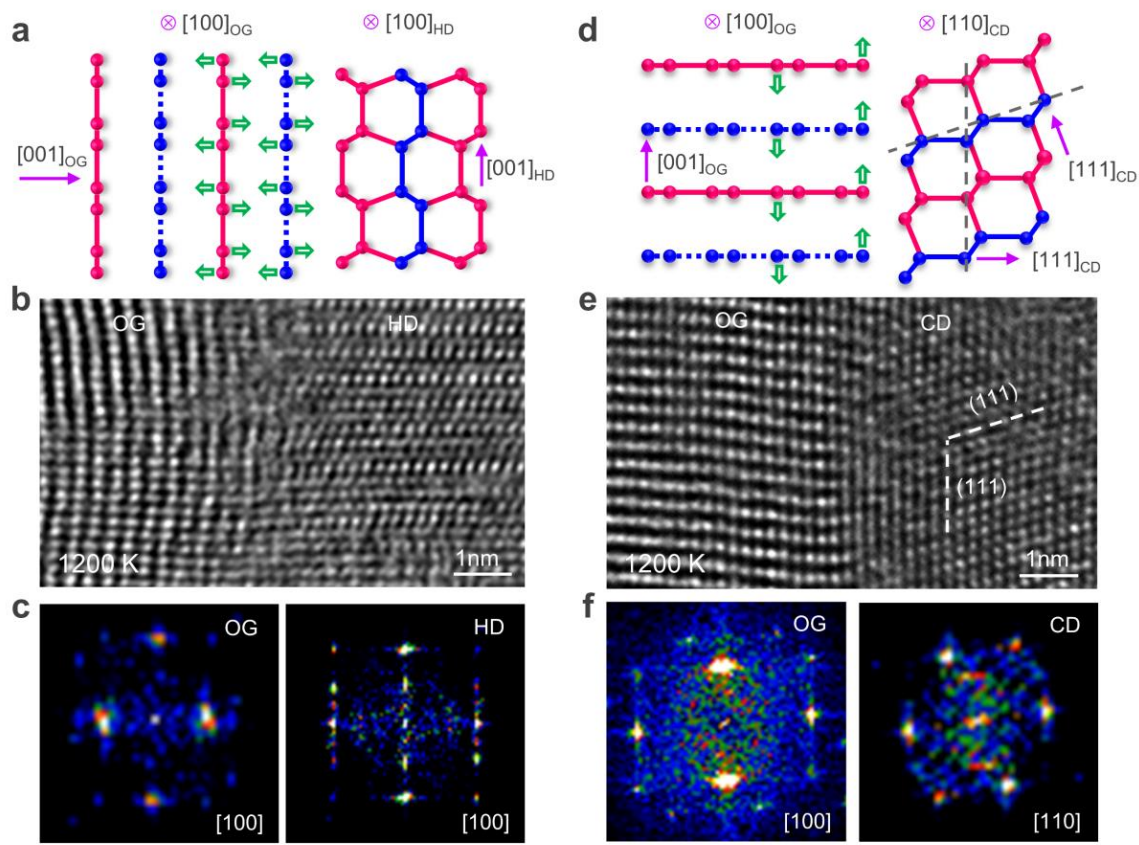

**Fig. 2. Atomistic observation of the direct G-D transformation pathway.** Schematic diagrams showing the hypothesized atomic arrangements for **a** HG→OG→HD and **d** HG→OG→CD through boat- or chair-buckling, respectively. HRTEM images showing the interfaces for **b** OG/HD and **e** OG/CD. Corresponding FFT patterns for **c** OG/HD and **f** OG/CD interfaces. Note the direct transformation from HG to HD is dominantly observed at the mild HPHT conditions.

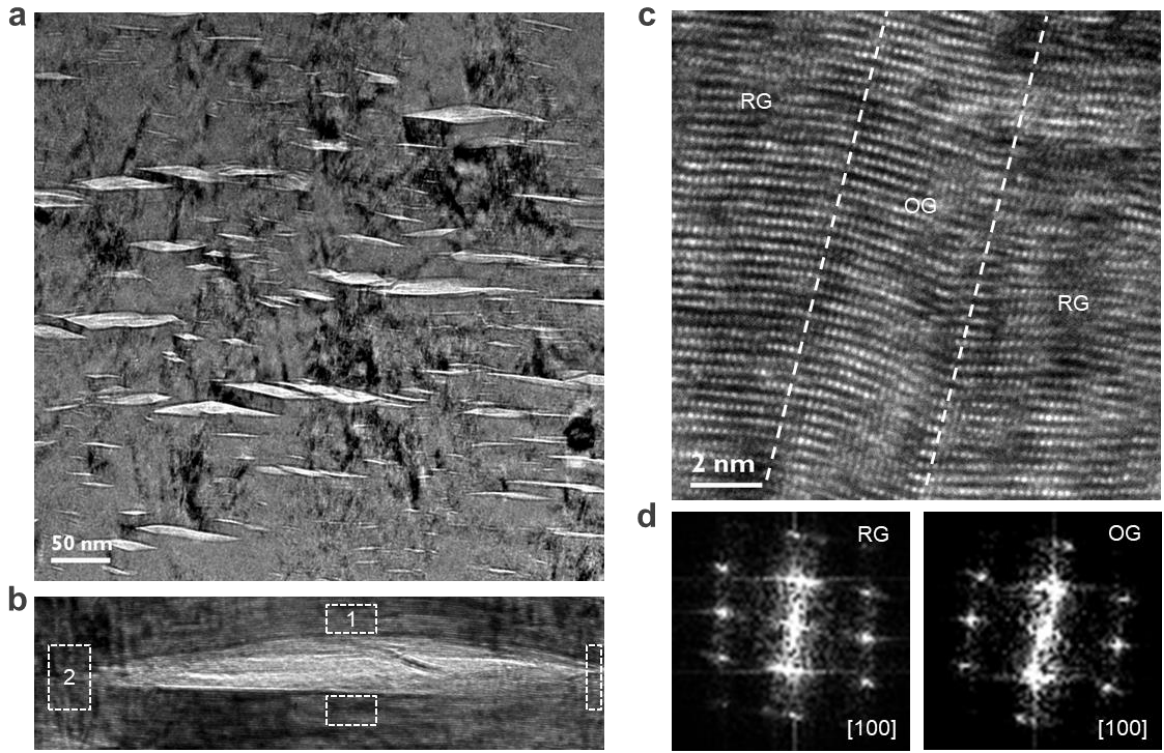

**Fig. 3. Intermediate OG phase stabilized in buckled graphite recovered from HPHT at 1000 K. a**, a low magnification TEM image showing rhombus voids. **b**, a higher magnification TEM image showing buckled areas marked by white boxes. **c**, HRTEM image of OG stabilized between two RG and its corresponding FFT in **d**.

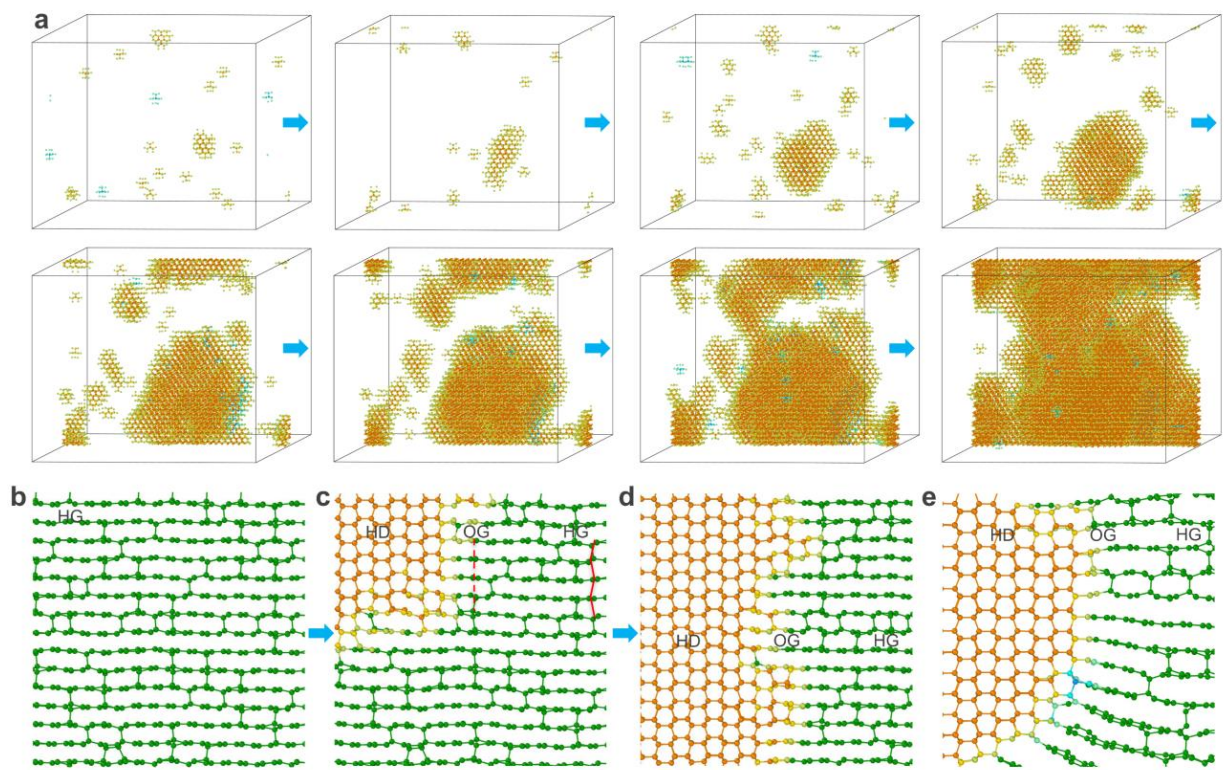

**Fig. 4. Simulated nucleation-growth mechanism from HG to HD. a**, 3D snapshots of the nucleation and growth of HD in graphite matrix. **b-d**, 2D snapshots of transformation pathway for HG→OG→HD. HG has slipped into OG phase and then transforms into HD. **e**, MD simulation of HG→HD transformation after pressure release, showing OG mediates between HG and HD.

## Methods

### Sample synthesis

Several single crystal graphite pieces have been subjected to high pressure and high temperature treatment in a diamond anvil cell (DAC) to synthesize the intermediate phases between HG and HD/CD. The base material is a 60×20 micron single crystal graphite disk cut from a millimeter size crystal by using the micro laser drilling system. Then, it was loaded into a hundred micron diameter rhenium gasket chamber. Pressure was monitored by ruby fluorescence. When the pressure raised to above 20 GPa, samples were heated to 1000 K, 1200 K and 1400 K by an YAG laser and recovered to ambient conditions.

### Sample preparation and structure characterization

We used Argonne Chromatic Aberration-corrected TEM (ACAT, a FEI Titan 80-300ST TEM/STEM) with a field-emission gun to investigate the crystallographic orientation, high-resolution transmission electron microscopy (HRTEM) images and electron energy loss spectra (EELS) of the recovered samples. The ACAT is equipped with a CEOS spherical and chromatic aberration imaging corrector to allow a resolution better than 0.08 nm information limit. To prepare the TEM sample, we first retrieved the quenched samples from high pressure and temperature treatments and transferred the sample from the DAC chamber to a clean marble mortar with a tiny pin. Then, we crushed the recovered sample into powder by using the marble mortar and pestle. Finally, we dispersed the crushed powder onto a holey carbon grid. We also used focused-ion beam (FIB) technique to prepare plane-view and cross-sectional TEM specimens.

### MD and DFT Simulations

**MD simulation.** The open-source MD code, LAMMPS[39] was used with LCBOP potential[40] for carbon to investigate the nucleation process for the direct graphite-to-diamond transformation. The initial hexagonal graphite model size is 13.2nm×10.2nm×13.6nm in the x, y and z directions, which are parallel to the [210], [100] and [001], respectively (containing 192,000 atoms). The interlayer distance of hexagonal graphite was first compressed to a certain value (0.208nm), Then model was equilibrated using the isobaric-isothermal (NPT) ensemble at certain temperature (2000 K) for 20 ps; during the equilibrium the interlayer distance of hexagonal graphite hold constant while the pressure of the other two directions were relaxed to zero. Finally, the relaxed model was compressed along the x direction with a constant strain rate $3.8×10^9$ s$^{-1}$. During the relaxation and compression process, the periodic boundaries conditions were applied in y and z directions. The simulation results were visualized and analyzed by the Open Visualization Tool (OVITO)[41].

**DFT calculation.** The enthalpy barrier for the HG to HD transformation were computed using the climbing image generalized solid state nudged elastic band (G-SSNEB)[42] calculations in VASP package[43] wherein the valence electrons are represented by plane wave basis set and the core electrons by the projector augmented wave method. We used the Perdew, Burke, Ernzerhof approximation[44] for the exchange-correlation energy. The convergence criteria for the total energy and forces per atom were set to $10^{-6}$ eV and 0.1 eV/Å respectively. The initial atomic positions for the intermediate images for the G-SSNEB were approximated by linear interpolation between the atomic positions of hexagonal graphite and hexagonal diamond.

**Acknowledgments**

This work was performed at the Center for Nanoscale Materials, a U.S. Department of Energy Office of Science User Facility, and supported by the U.S. Department of Energy, Office of Science, under Contract No. DE-AC02-06CH11357. We thank the Jie Wang's help on the sample preparation. D. L. thanks the China Scholarships Council (CSC) Joint PhD Training Program for the financial support of studying abroad.

**Author contributions**

D.L. and J.G.W. performed the TEM characterization. L.X.Y., W.G.Y. and H.K.M. synthesized the high-pressure and high-temperature samples. D.L and J.G.W. prepared the TEM samples. D.L., J.G.W. processed and analyzed the data and discussed with J.S.T., L.X.Y., W.G.Y., Su.S., N.R., I.A., and H.K.M. H.X.X. performed the MD calculations. Sr.S. and Su.S. performed the DFT calculations. D.L., J.G.W. wrote the manuscript with discussion and improvements from all authors.

**Supplementary Figures**

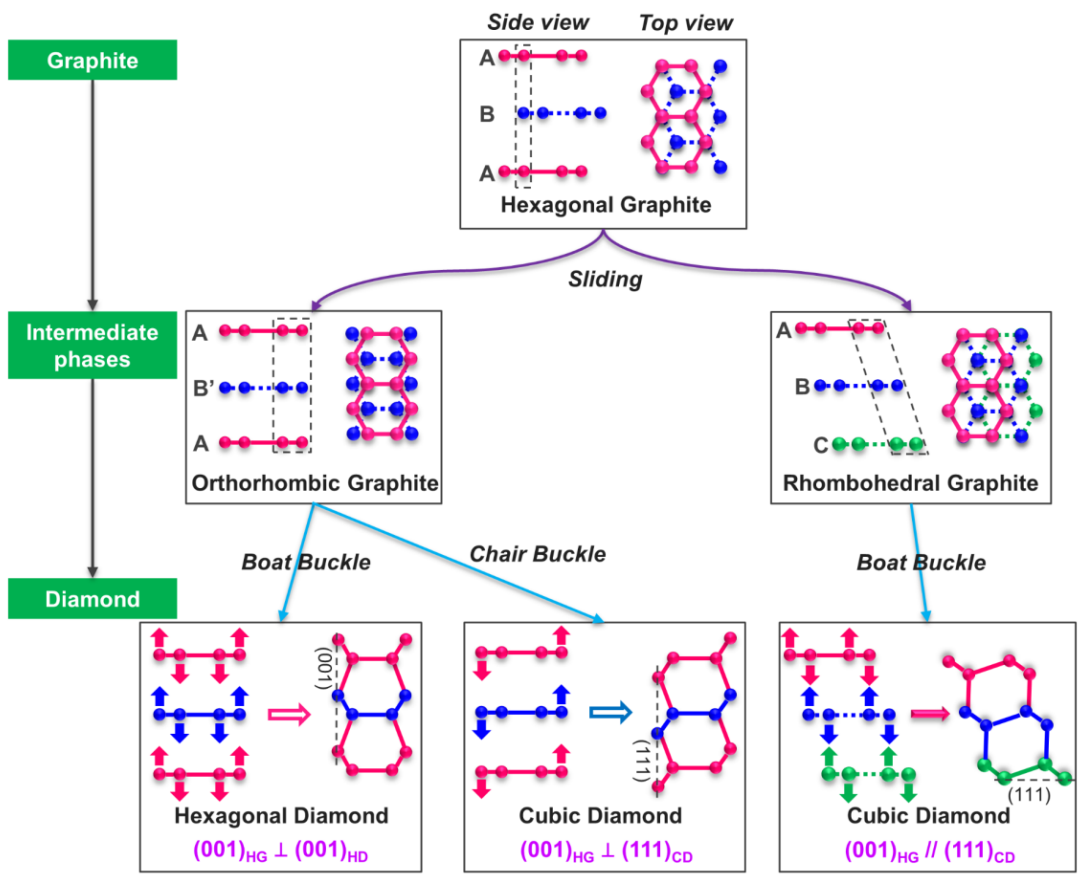

**Fig. S1. Possible pathways for hexagonal graphite-to-diamond transition.**

In hexagonal graphite, carbon atoms are arranged in planar honeycomb sheets with AB stacking, which hinders direct transformation into diamond due to the lack of matching carbon atoms at honeycomb centers with carbon atoms in adjacent layers. Therefore, extensive theoretical research suggests that the direct transformation of AB-stacked hexagonal graphite (HG) to hexagonal diamond (HD) or cubic diamond (CD) must first undergo a sliding process to form an intermediate phase, including either orthorhombic graphite (OG) with AB' stacking or rhombohedral graphite (RG) with ABC stacking, followed by boat or chair buckling to form $sp^3$ bonds. When the transformation goes through OG, (001) honeycomb planes in HG is perpendicular to corrugated densest planes (001) planes in HD and (111) in CD. In case of RG, (001) honeycomb planes in HG is parallel to (111) in CD.

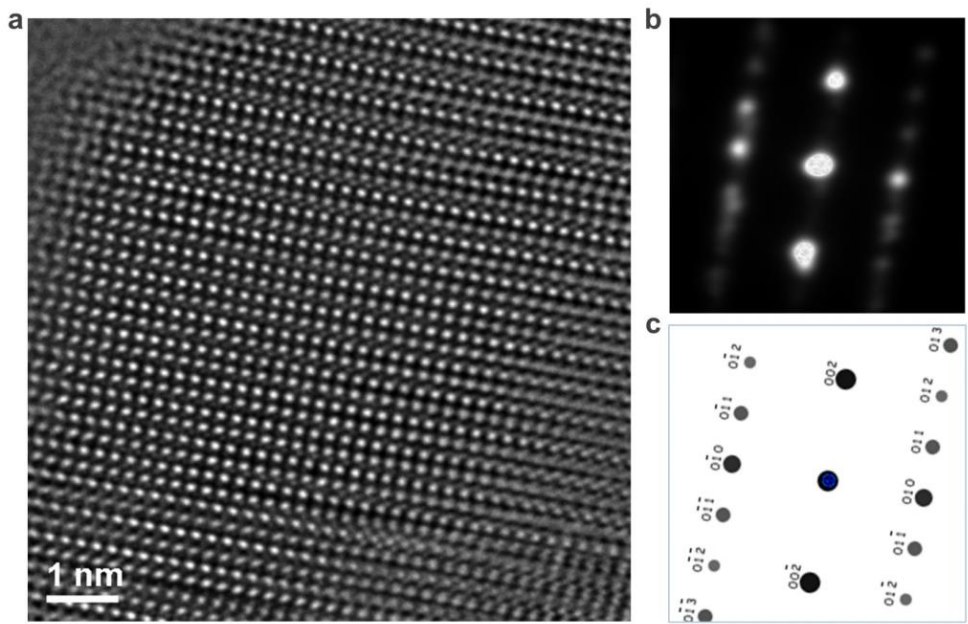

**Fig. S2. High-quality single-phase HD from recovered HPHT at 1400 K samples. a**, TEM image of [100]; **b**, Selected area electron diffraction (SAED) of $[100]_{HD}$; **c**, Simulated electron diffraction of $[100]_{HD}$., which matches with the diffraction pattern in **b**.

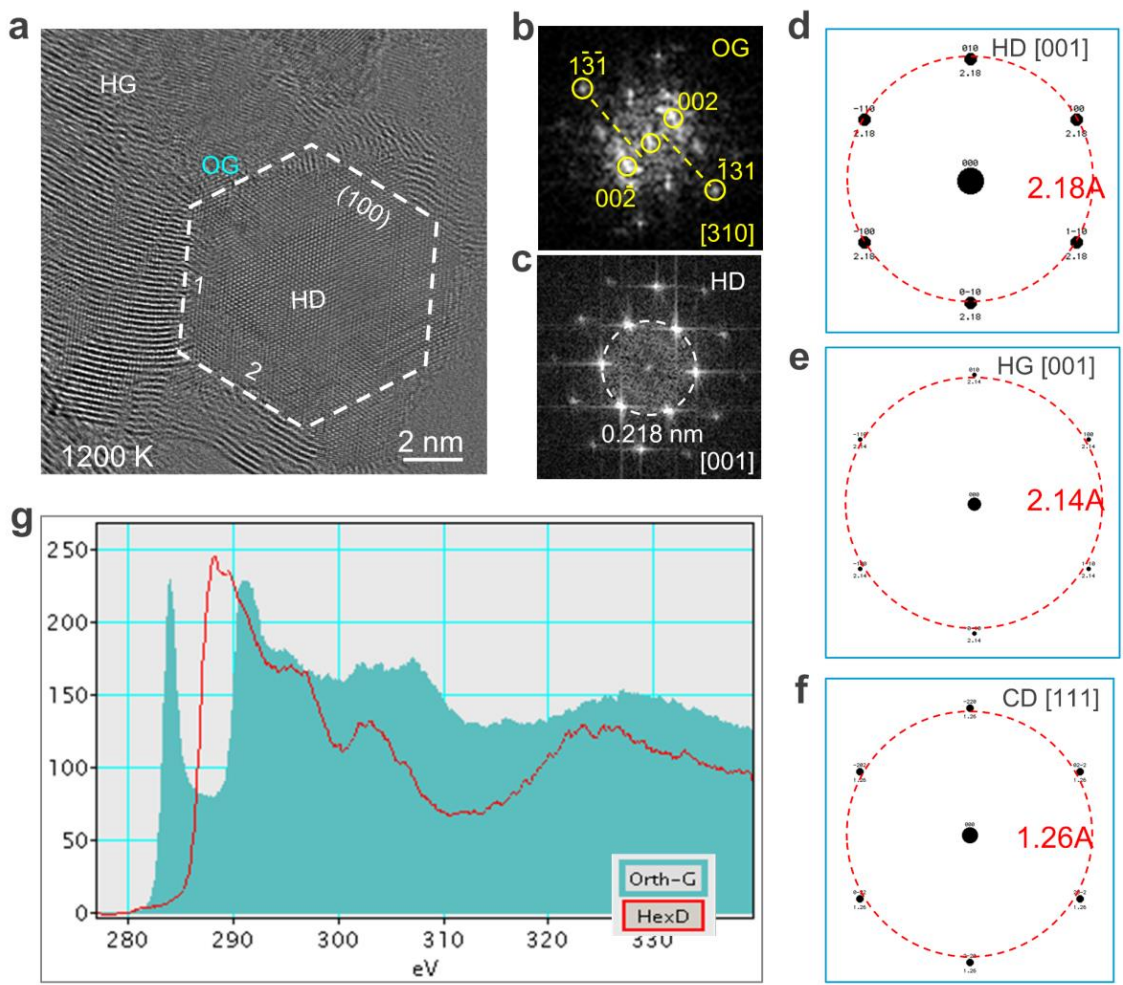

**Fig. S3**. 6-fold symmetry diffraction pattern. **a**, HRTEM image of nanoscale HD nuclei embedded in the graphite matrix. **b**, The FFT pattern from the OG in **a** shows a diffraction pattern matching with $[310]_{OG}$ instead of $[1\bar{1}0]_{HG}$. **c**, the FFT pattern from the HD in **a** has a 6-fold symmetry with a 0.218 nm lattice spacing indexed by $[001]_{HD}$. **d**, [001] diffraction pattern of a HD structure; **e**, [001] diffraction pattern of a HG structure; **f**, [111] diffraction pattern of a CD structure. **g**, EELS spectra of HD and distorted graphite OG near the nanodiamond.

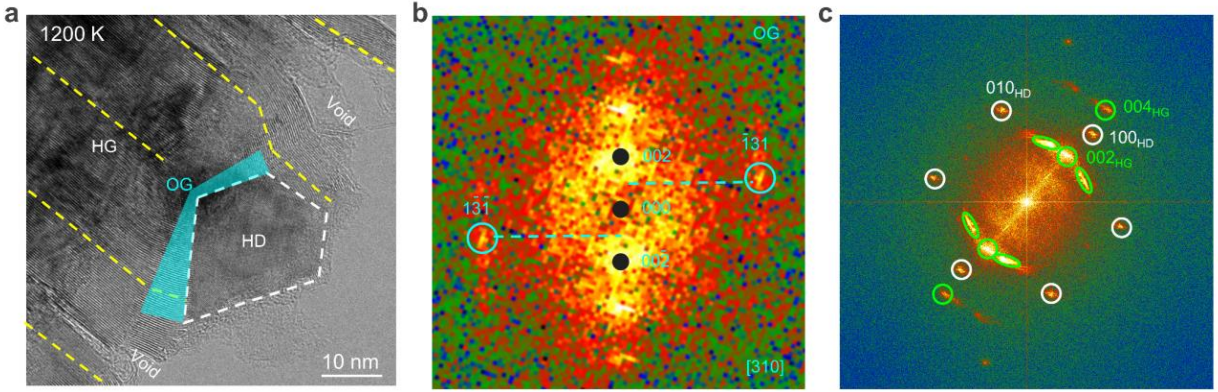

**Fig. S4. Interfaces between nono-diamond and graphite matrix. a**, TEM image of graphite-diamond interfaces; **b**, FFT pattern from the OG area. Note the first-order diffraction spots lie in the middle of the zero-order diffraction spots. In contrast (also see Fig. S5f), the first-order diffraction spots are aligned with the zero-order diffraction spots in the case of HG; **c**, FFT pattern from the HD nanodiamond area, showing that the $(002)_{HG}$ is more or less parallel to $(001)_{HD}$ (~4 degree off). Two $(002)_{HG}$ arcs indicated by yellow arrow heads reflect the bent graphitic planes where nanodiamond are found.

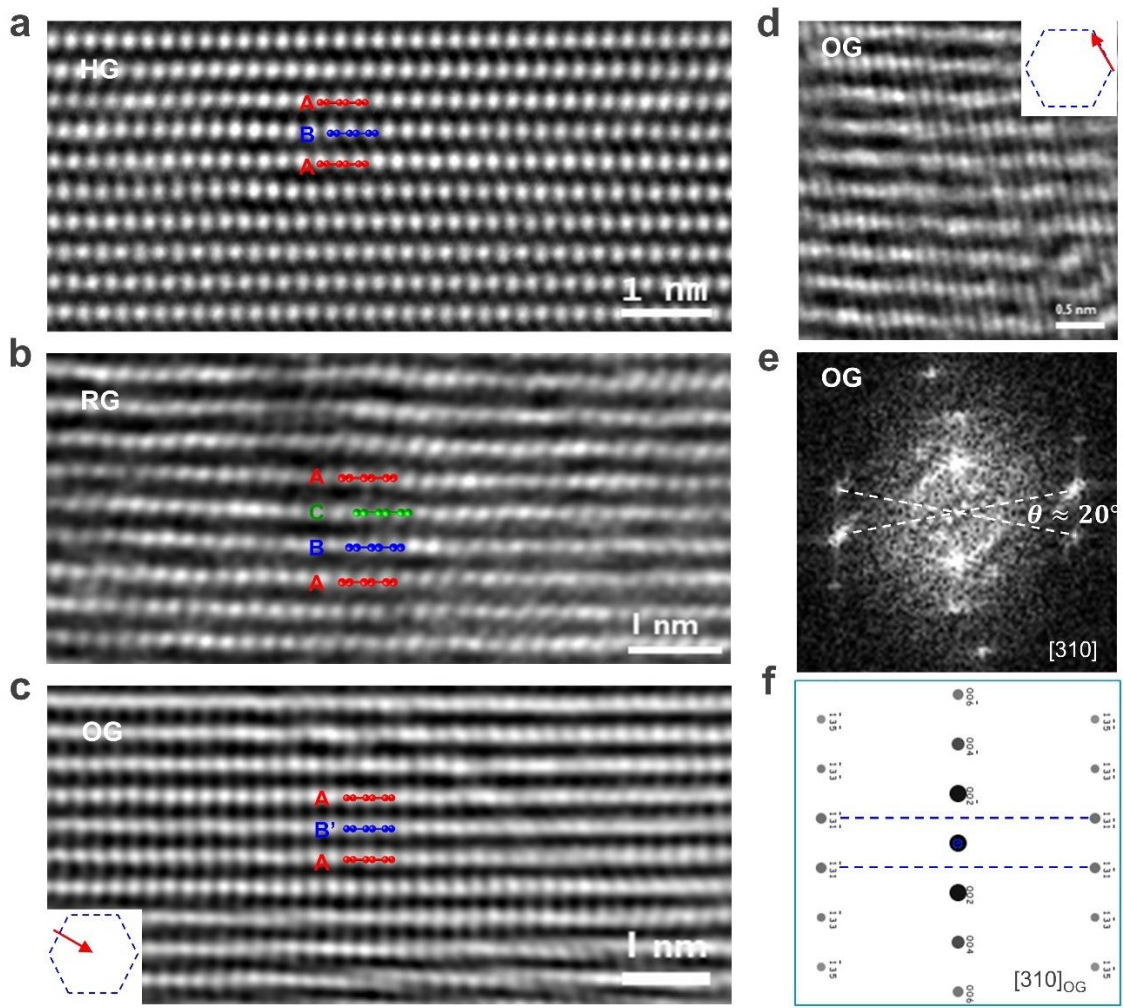

**Fig. S5. High-resolution TEM images of graphite structures. a**, HG with AB stacking; **b**, RG with ABC stacking; **c**, OG with AB' stacking; **d-e**, HRTEM image and [310] diffraction of OG show AB' stacking instead of AA stacking.

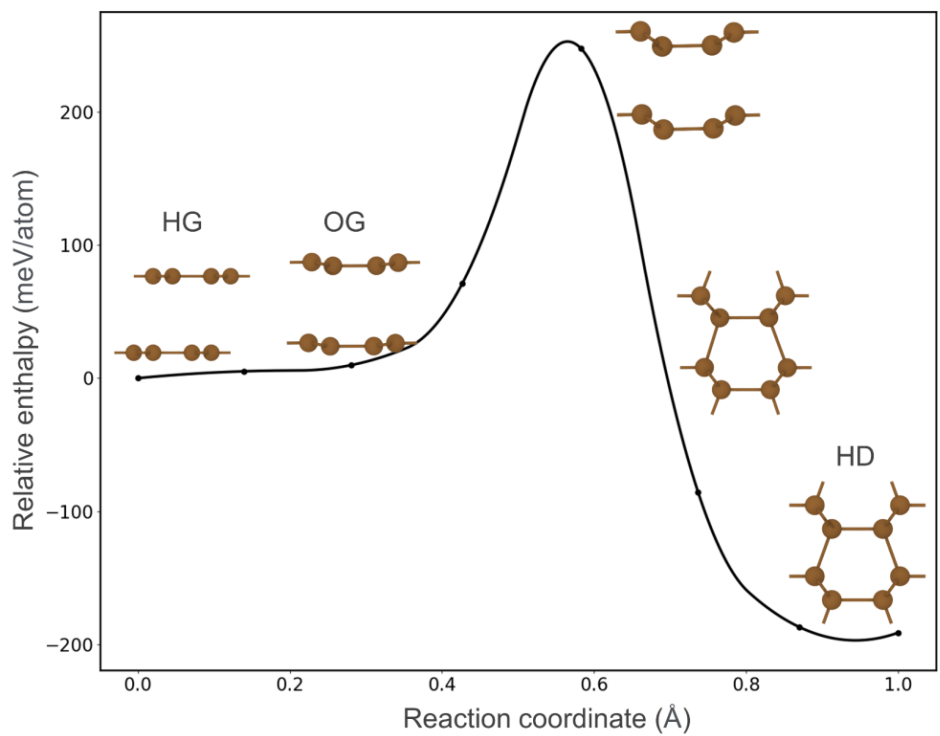

**Fig. S6. The enthalpy landscape and pathway for concerted HG→OG→HD transformation.**